\documentclass[10pt,a4paper]{article}
	\usepackage[utf8]{inputenc}
	\usepackage{amsmath}
	\usepackage{xcolor} 
	\usepackage{hyperref}
	\usepackage{amsfonts}
	\usepackage{amssymb}
	\usepackage{fancyhdr}
	\usepackage{graphicx}
	\usepackage{eurosym}
	\hypersetup{colorlinks=false, urlcolor=black, allbordercolors=white,}
	\author{Giovanni Bindi (\textit{giovanni.bindi@istruzione.it})\\
	 Istituto Zoli, Atri, TE, Italy}
	\title{Improving physics with CanSat}
	\date{}
	
\begin{document}
	\lhead{}
	\chead{}
	\rhead{}
	\lfoot{}
	\cfoot{}
	\rfoot{\thepage}
	\maketitle
	\begin{abstract}
	The european CanSat competition is a funny way to understand physics. My students used low budget sensors and data analysis testing a model for temperature and humidity in low troposphere.
	\end{abstract}\newpage\
	\linebreak 
	\begin{figure}
	\begin{center}
\includegraphics[scale=.7]{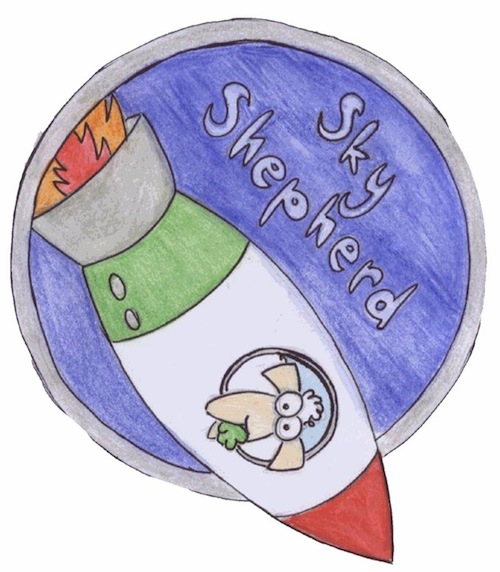}
\caption{The logo : a sheep inside a rocket}
\end{center}
\end{figure}
	\section{The CanSat}
My high school students, two girls and four boys, seventeen years old, have been chosen by ESA to participate to CanSat 2015 European competition (16 teams from all the Europe). A CanSat is a small satellite inside a soft drink can, launched by a rocket to an altitude of about 1000 m with an acceleration of 10$g$. The CanSat primary mission is measuring pressure and temperature, sending data to ground station and landing safely. Secondary mission consists of a scientific experiment different for each team.
	The 2015 European CanSat Competition was organized by the European Space Agency (ESA) in collaboration with AeroEspaço – Air and Space Science Center. The finished CanSats were launched from the Santa Cruz Air Field, 60 km north of Lisbon, Portugal, on 24-28 June 2015.
	The European CanSat competition is part of ESA’s initiative to inspire young people to follow a career in science or engineering, with a view to ensuring the availability of a highly qualified work force in the space industry of the future.
	The team logo of my students features a handmade sheep inside a rocket up in the skies, that perfectly represents the team name 'Sky Shepherd' and also the roots of our region, Abruzzo, that has always been a land of shepherds and farmers.
\subsection{Sensors}
\begin{figure}
\includegraphics[scale=0.29]{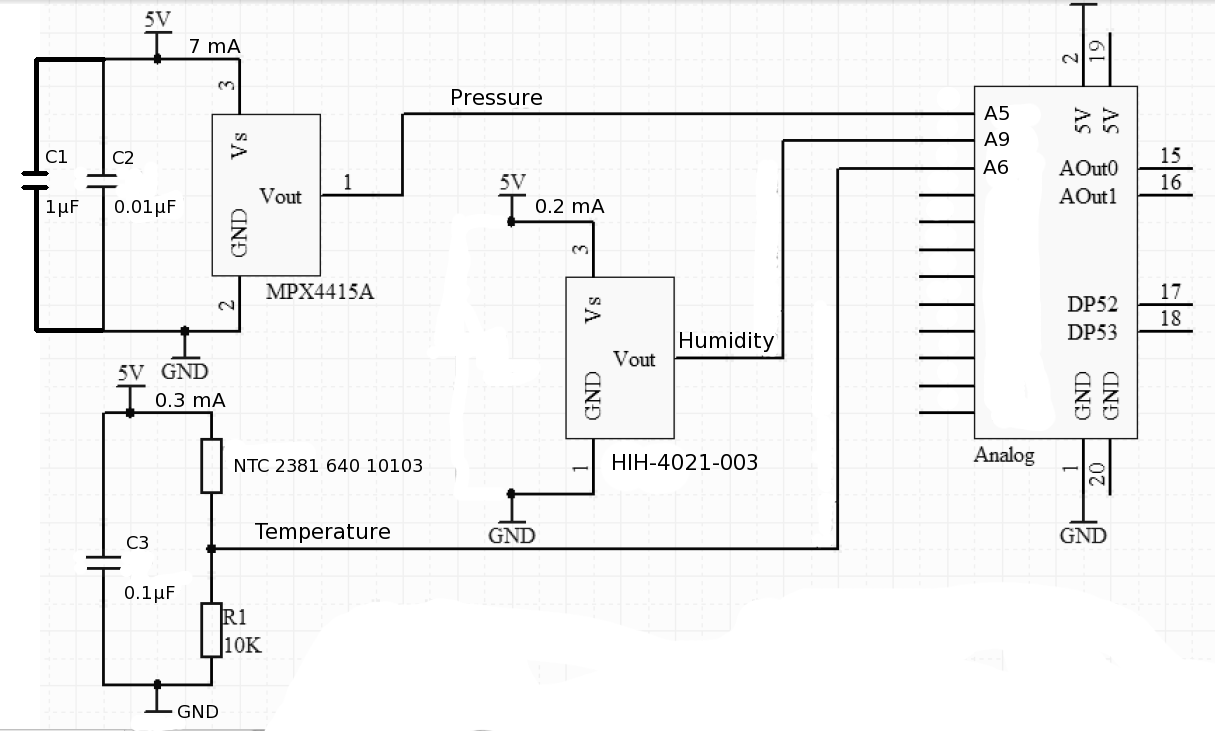}
\caption{Sensors schematic}\label{schema}
\end{figure}

The students placed inside the Cansat three sensors connected to Arduino analog inputs. There are some capacitors for filtering and power supply decoupling. Sensors schematic is in Figure \ref{schema}.\\
For the primary mission they used the MPX4115 pressure sensor included in the CanSat kit from ESA, because it is designed to sense absolute air pressure in an altimeter applications. Typical current is $7.0 mA$ at $5.0 V$.\\
The temperature sensor was a NTC thermistor which has excellent accuracy between 25 $^{o}C$ and 85 $^{o}C$ and high stability over a long life. Typical current is $0.3 mA$ at $5.0 V$ with load resistor $10 K\Omega$.\\
For the secondary mission students chose the humidity sensor Honeywell HIH-4021-003 with a specific calibration and data printout. This sensor is ideally suited for low drain, battery operated systems.
Sensor is ratiometric to supply voltage and light sensitive. For best performance, it is important to shield sensor from bright light. Typical current is $0.2 mA$ at $5.0 V$.
\subsection{Interface Arduino-Data analysis}
Regarding the interface Arduino - data analysis they used the program Cool Term. Cool term is a simple serial port terminal application that is used by amateurs and professionals with a need to exchange data with hardware connected to serial ports such as servo controller, robotic kits, gps receivers , microcontrollers ( such as Arduino ). By means of this program pupils saved data in real time as text file from the Arduino's serial monitor.
\subsection{Temperature-altitude theory}
An interesting application \textbf{\cite{fermi}} of the adiabatic gas expansion is the dependence calculation of the atmosphere temperature on the height above the sea level.\\
When air rises from sea level to the upper region of lower pressure, it expands. Since air is a poor conductor of heat we may consider the expansion as taking place adiabatically, so the temperature of the rising air decreases.\\
We can consider a column of air where $dp$ is the pressure variation, $\rho$ is the air density, $g$ is the gravity acceleration, $dh$ is the height variation\\
	in formula we have \begin{equation}
	dp =-\rho gdh
	\end{equation}
	From the gas law
	\begin{equation}
	pV=nRT
	\end{equation}
	where $n = \dfrac{m}{M}$, with $M$ the molar mass of air and $m$ the total mass of air\\
	\\
	using $\rho = \dfrac{m}{V}$
	we can write
	\begin{equation}
	\rho = \dfrac{pM}{RT}
	\end{equation}
	\begin{equation}
	dp=-\dfrac{pM}{RT}gdh
	\end{equation}
	Using the adiabatic formula
	\begin{equation}
	T \propto p^{\dfrac{\gamma-1}{\gamma}}
	\end{equation}
	we can take the logarithmic derivative
	\begin{equation}
	\dfrac{dT}{T}=\dfrac{\gamma-1}{\gamma}\dfrac{dp}{p}
	\end{equation}
	from the (4) we have
	\begin{equation}
	\dfrac{dp}{p}=-\dfrac{M}{RT}gdh
	\end{equation}
	so we can combine (6) and (7)
	\begin{equation}
	\dfrac{dT}{T}=-\dfrac{\gamma-1}{\gamma}\dfrac{M}{RT}gdh
	\end{equation}
	the final formula is
	\begin{equation}
	\dfrac{dT}{dh}=-\dfrac{\gamma-1}{\gamma}\dfrac{Mg}{R}
	\end{equation}
	Assuming $\gamma=\dfrac{7}{5}$,	\hspace{1cm}$g=9.81 \dfrac{m}{s^{2}}$, \hspace{1cm}$R=8.31 \dfrac{J}{K}$
	\hspace{1cm}$M=29.0 \dfrac{g}{mol}$\\
	we have \begin{center}
	$\dfrac{dT}{dh}=-9.8\dfrac{^{o}C}{1000m}$
	\end{center}
	\vspace{.5cm}
	This number is larger than the observed value, in fact the international standard value is
	\begin{center}
	$\dfrac{dT}{dh}=-6.5\dfrac{^{o}C}{1000m}$
	\end{center}
	 A possible reason for the difference is that we have neglected the effect of water vapor condensation. 
	\subsection{Pressure-altitude theory}
	Integrating the (7) yields\begin{equation}
	ln(p)=-\dfrac{M}{RT}gh+C
	\end{equation}
	using the boundary condition yields
	\begin{equation}
	h=-\dfrac{RT}{Mg}ln(\dfrac{p}{p_{0}})
	\end{equation}
	where $p_{0}$ is the base pressure, i.e. the pressure at the ground level.
	This is the hypsometric equation but it assumes that temperature remains constant, using :\\

$T=288 K$ \hspace{1cm} $M=29.0 \dfrac{g}{mol}$ \hspace{1cm} $R=8.31\dfrac{J}{K}$,\\
	  
$p_{0}=101300Pa$ \hspace{1cm} $p=96500Pa$ \\

	   we can calculate \begin{center}$h=409m$\end{center}	In order to take the temperature as a variable we need to introduce the lapse rate $L$. The lapse rate is defined as the rate of temperature increase in the atmosphere with increasing altitude.\\
	The standard lapse rate near the ground is assumed to be $L=-6.5 \dfrac{^{o}C}{1000m}$\\
	Assuming a constant lapse rate, a better relation is\begin{equation}
	h=\dfrac{T_{0}}{L}(\dfrac{p}{p_{0}}^{-\dfrac{LR}{Mg}}-1)
	\end{equation}
	where $p_{0}$ is pressure at ground level and $T_{0}$ is temperature at ground level. Using for the (12) the same values of (11) with the standard lapse rate, we can calculate \begin{center}$h=407m$\end{center} The final results for (11) and (12) are very similar.\\
	
An empirical formula for low troposphere is the international standard value
\begin{center}
$\dfrac{dp}{dh}=-11.15\dfrac{hPa}{100m}$
\end{center}
	\subsection{Parachute}
Students bought different parachutes so they decided to choose the best one by calculating their speeds with a constant mass. In fact, they designed an experiment measuring velocity for each parachute, knowing all the other parameters.\\

In this experiment, the pupils decided to launch each parachute from a fixed altitude $h$ together with a constant mass, $m_{object}$, a can of coke with a mass of about 350 g. Altitude \textit{h} was divided in 9 steps of constant height $x=3.30m$, i.e. the height of each storey.\\

The scholars analysed the trajectory of fall (YouTube video: \url{https://www.youtube.com/watch?v=YNOonlnorhM} ) of the parachute-object system, they calculated the speed for each step and the arithmetic mean between the values obtained.\\

\begin{figure}
\begin{center}
\includegraphics[scale=0.8]{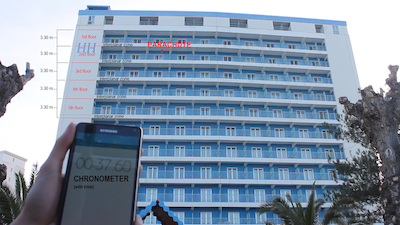}
\end{center}
\caption{A view of the hotel and the experiment.}\label{snap}
\end{figure}

The place they chose, thanks to the availability of the manager, was the hotel “Hermitage” placed at sea level, in the town of Silvi Marina (TE). Students conducted the experiment on 10th April 2015, with a wind blowing towards north-west with an intensity of 11.1 km/h. They launched the parachutes from the twelfth floor of the hotel, from a height between 42.70 m and 45.70 m.(Figure~\ref{snap})\\

Each parachute+object system was thrown in such a way to limit as much as possible the contribution of the speed during the first part of the descent motion in which the system did not proceed with a constant speed (acceleration decreases progressively until the system reaches a steady state in which it moves at constant speed). Trying to avoid the parallax error students filmed very far from the hotel and they did several launches using two different parachutes.\\
\subsubsection{Experiment with PAR-18 Thin}
\begin{figure}[htbp]
\centering
\includegraphics[scale=0.5]{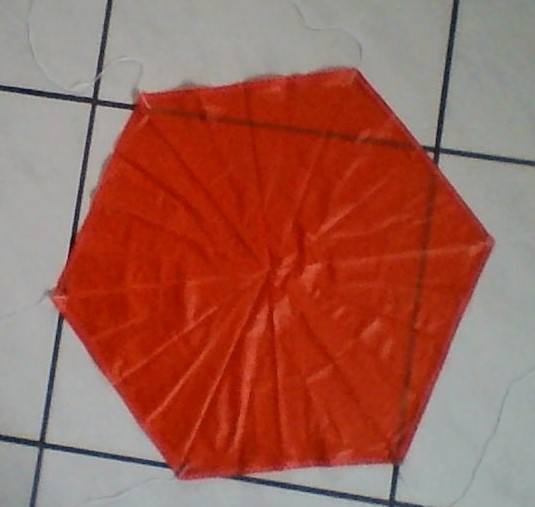}
\caption{PAR-18 Thin – Parachute Top Flight}\label{PAR18}
\end{figure}
Several launches were performed with the PAR-18 Thin – Parachute Top Flight\\ (Figure~\ref{PAR18}). Its technical features are:\\
\begin{itemize}
\item \textit{Material:} ripstop nylon with nylon cords sewn;
\item \textit{Diameter:} 52 cm;
\item \textit{Shape:} hexagonal;
\item \textit{For mass:} 226-450 g;
\end{itemize}
The table below represents the results of the first launch.
\begin{figure}[htbp]
\centering
\begin{center}
\includegraphics[scale=0.70]{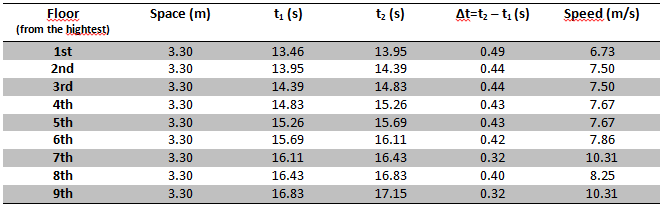}
\end{center}
\end{figure}\\
Velocity of the parachute+object system using the PAR-18 Thin with a $m_{object}$ of 350 g  is:
\begin{center}
$\overline{v_{m}}=(8.20\pm1.25)m/s$
\end{center}
\begin{figure}[htbp]
\centering
\includegraphics[scale=0.30]{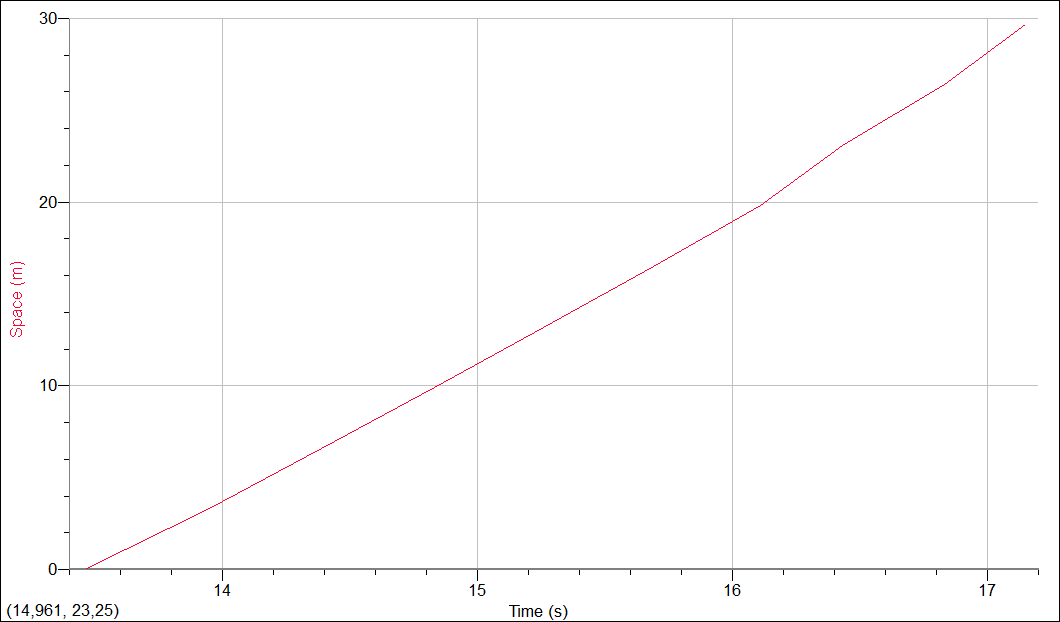}
\caption{PAR-18 First launch - space vs. time}\label{Grafico001}
\end{figure}
Looking at the space-time plot of the first launch (Figure~\ref{Grafico001}), vertical motion of the
parachute+object system seems very similar to a uniform motion.\\
\subsubsection{Experiment with XTPAR-24 Thin}
\begin{figure}[htbp]
\centering
\includegraphics[scale=0.3]{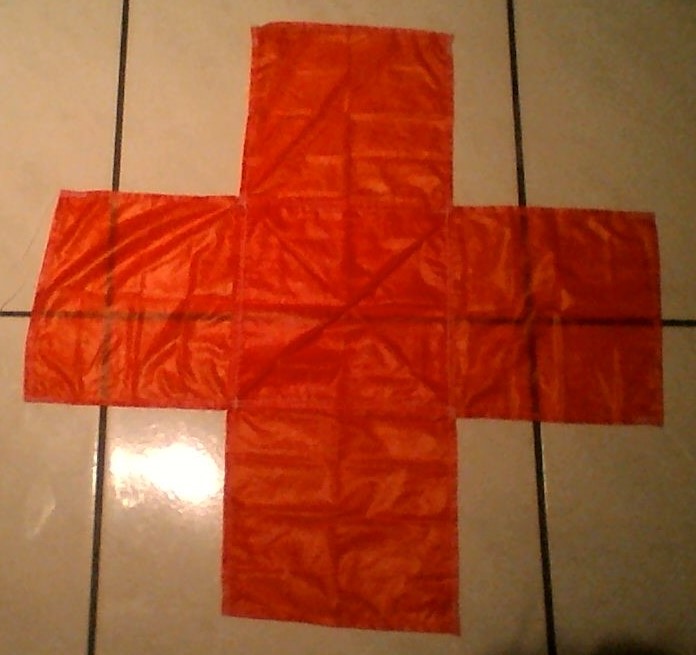}
\caption{XTPAR-24 Thin – Parachute Top Flight}\label{XTPAR-24}
\end{figure}

\begin{figure}[htbp]
\centering
\includegraphics[scale=0.30]{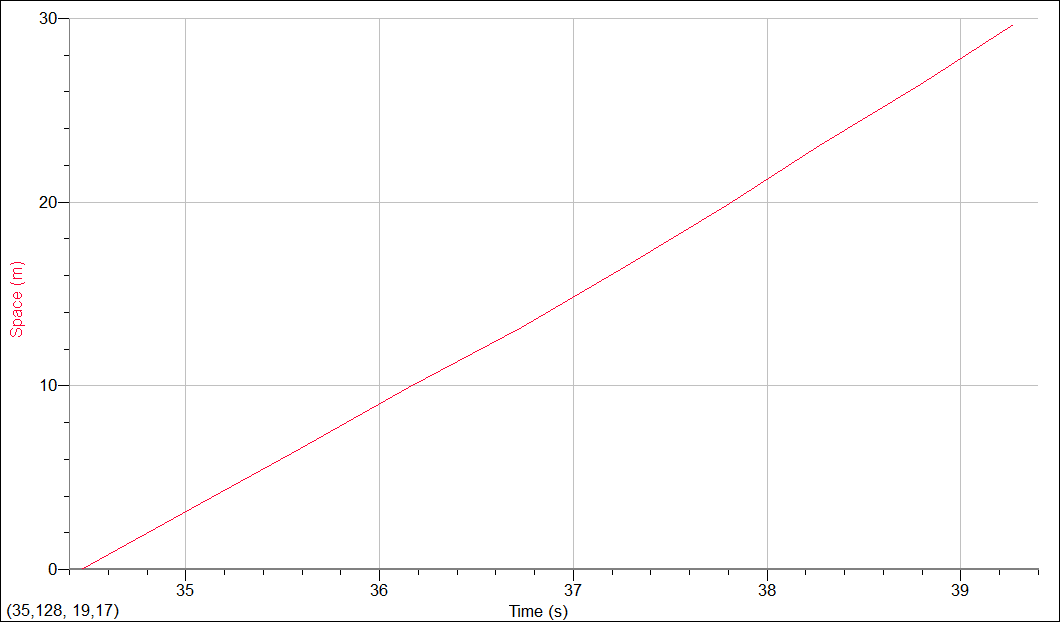}
\caption{XTPAR-24 First launch - space vs. time}\label{Grafico003}
\end{figure}

Several launches were done with the XTPAR-24 Thin – Parachute Top Flight.\\ (Figure~\ref{XTPAR-24}) Its technical features are:
\begin{itemize}
\item \textit{Material:} nylon fabric with ripstop nylon cords sewn;
\item \textit{Diameter:} 61 cm;
\item \textit{Shape:} cruciform parachutes;
\item \textit{For mass:} 340-450 g;
\end{itemize}
The space-time plot of the XTPAR-24 first launch (Figure~\ref{Grafico003}) shows that the vertical motion of the parachute+object system was very similar to a uniform motion.
\begin{center}
\includegraphics[scale=0.70]{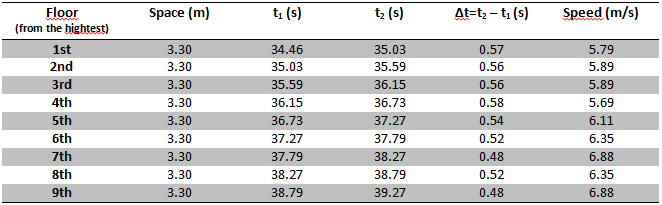}
\end{center}
This parachute had a higher stability and suffered less influence from external factors than the first, as the system has maintained values of speed more constant than the first parachute.
The standard deviation of the second parachute $\sigma=0.75m/s$ is lower than the first one $\sigma=1.25m/s$. This fact reinforces the hypothesis that the second parachute is more stable than the first. The stability of a parachute is the tendency to return to the center position when subjected to oscillations and this allows the maintenance of a fairly constant descent speed. Indeed the most stable parachutes are the cruciform ones that oscillate within a span of very few degrees.\\
Steady velocity of the parachute+object system using the XTPAR-24 Thin with a $m_{object}$ of 350 g is:
\begin{center}
$\overline{v_{m}}=(6.15\pm0.75)m/s$
\end{center}
The experiment proved that the second parachute not only had a lesser descent speed but also a better stability and for these reasons has been chosen by the pupils.\\
\subsection{Antenna}
Students built a “Yagi antenna” able to receive a signal of 433.35 MHz from the CanSat. It is made up of several elements, mutually parallel, assembled on a support called 'boom'. Antenna's directivity provides good attenuation against signals from directions other than pointing.
\begin{figure}
\centering
\includegraphics [scale=0.5]{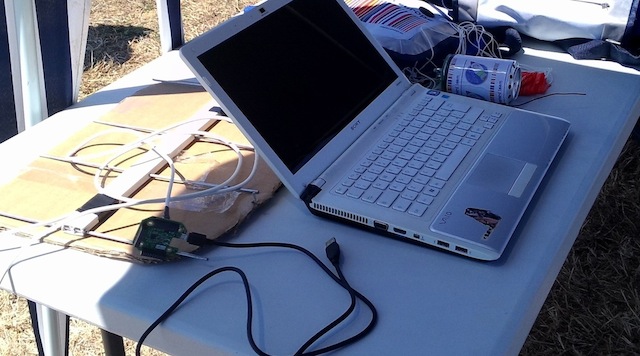}
\caption{The antenna with the transceiver and the CanSat just before the launch}
\end{figure}
Using makeshift materials, they realized a rudimentary Yagi antenna, in order to strengthen the signal of the transceiver. Firstly, the scholars built the ‘boom’ (made of insulating material) using a plastic support by cutting the sides, so as to give a length of 392 mm. With a drill, they realized the holes where to install the antenna components, made of aluminum. The pupils connected the cable to the axial dipole, which is going to receive the signal on one side, and they connected the grounding on the opposite side.\\
\begin{figure}
\centering
\includegraphics [scale=1.5]{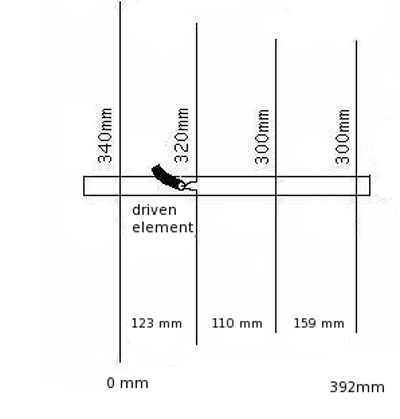}
\caption{Antenna design}\label{diagram}
\end{figure}
In theory to built an antenna Yagi the driven element must have a length of  $\lambda /2$   but it came out that the best length is 0.93 $\lambda /2$. In this case the driven element resulted $320 mm$ (Figure~\ref{diagram}).
 \section{Measurements}
\subsection{Calibration}
Preliminary step was sensors calibration using datasheet conversion tables and reference values in order to obtain a more precise measure of pressure, temperature and relative humidity. Students produced a calibration plot for each of the three sensors and using Logger Pro \textit{'curve fit function'} they produced three conversion formulas.
\begin{figure}
\centering
\includegraphics [scale=0.25]{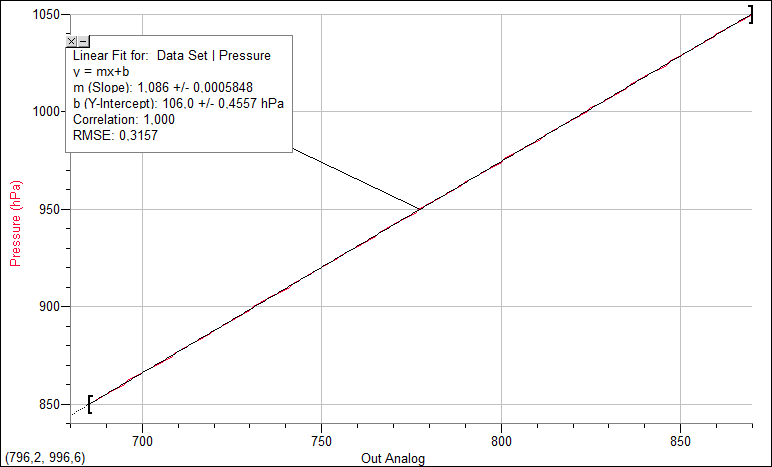}
\caption{Pressure calibration}\label{pressurefit}
\end{figure}

Pressure conversion formula obtained (Figure~\ref{pressurefit}) is:
\begin{center}
$P=1.086x+106.0$
\end{center}
Where $x$ is 10 bit Arduino reading (0 $\div$ 1023) of sensor analog voltage and $P$ is the value of pressure ($hPa$).

Temperature calibration is more interesting because the NTC sensor has not a linear response.
\begin{figure}
\centering
\includegraphics [scale=0.25]{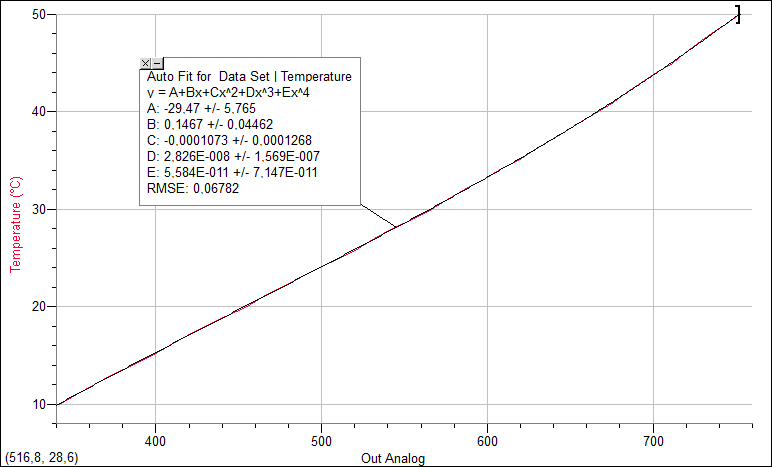}
\caption{Temperature calibration}\label{temperaturefit}
\end{figure}
The conversion formula of temperature they obtained (Figure~\ref{temperaturefit}) is a polynomial of degree four:
\begin{center}
$T=5.584\cdot10^{-11}x^{4}+2.826\cdot10^{-8}x^{3}+1.073\cdot10^{-4}x^{2}+1.467\cdot10^{-1}x+29.47$
\end{center}
Where $x$ is 10 bit Arduino reading (0 $\div$ 1023) of sensor analog voltage and $T$ is the value of temperature ($^{o}C$).
\begin{figure}
\centering
\includegraphics [scale=0.25]{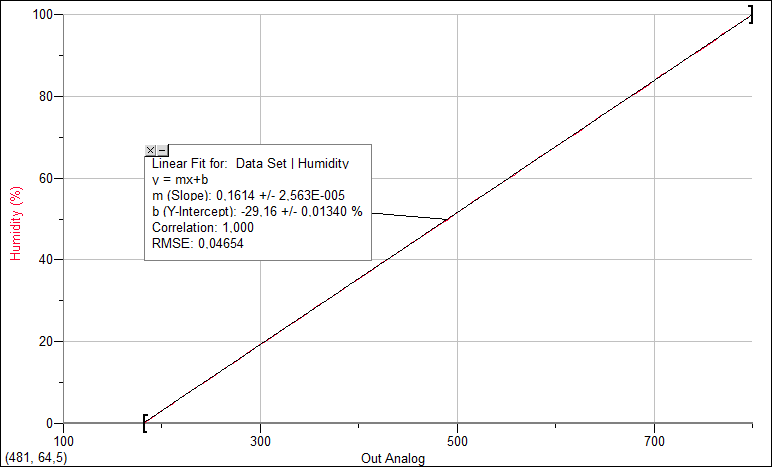}
\caption{Humidity calibration}\label{humidityfit}
\end{figure}

The conversion formula of relative humidity they obtained (Figure~\ref{humidityfit}) is:
\begin{center}
$RH=0.614x-29.16$
\end{center}
Where $x$ is 10 bit Arduino reading (0 $\div$ 1023) of sensor analog voltage and $RH$ is relative humidity.
 \subsection{Flight test data}
\begin{figure}
\centering
\includegraphics [scale=0.35]{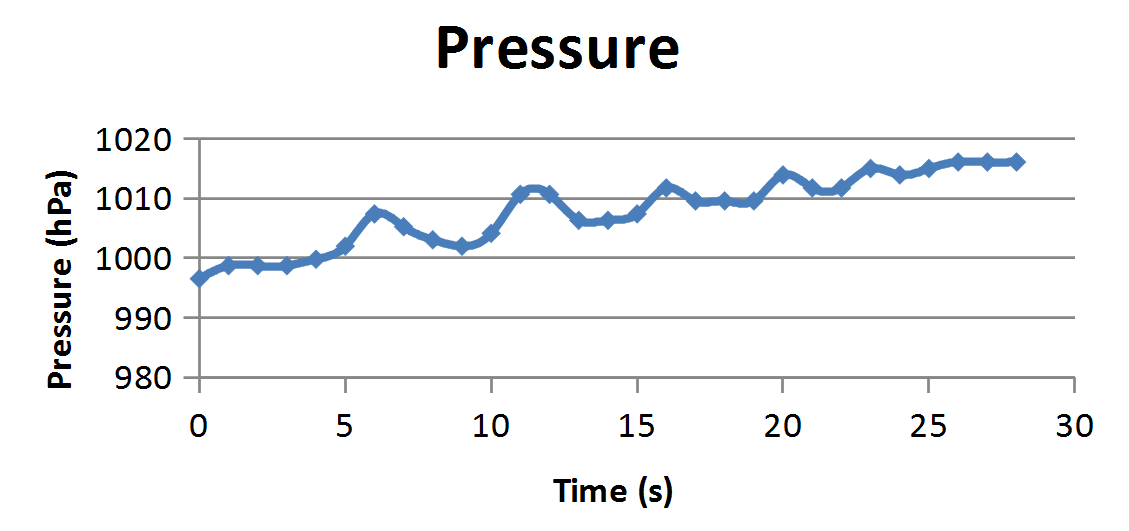}
\caption{Flight test : pressure vs time}\label{flightpressure}
\end{figure}
As a test, kids launched the CanSat from a Cessna plane at an altitude of 165 m.
Radio transmission was completely satisfactory because they got all data, before, during and after the launch of the CanSat.
\subsubsection{Pressure}
The pressure vs time diagram is shown in Figure~\ref{flightpressure}: pressure is calculated every second during the falling motion. (At time $t=0s$ the CanSat was thrown from the plane, while $t=28s$ is the landing time).
It was very simple to analyse data in this case as students had no problem in data reception. They measured the pressure at the ground and the pressure at the maximum height ($165m$ according to airplane altimeter).
Considering true the maximum height, the pressure variation with altitude they measured is:
\begin{center}
$\dfrac{dp}{dh}=-19.55\dfrac{hPa}{165m}$
\end{center}
that is:
\begin{center}
$\dfrac{dp}{dh}=-11.85\dfrac{hPa}{100m}$
\end{center}
This is very similar to the international standard in low troposphere:
\begin{center}
$\dfrac{dp}{dh}=-11.15\dfrac{hPa}{100m}$
\end{center}
\subsubsection{Temperature}
\begin{figure}
\centering
\includegraphics [scale=0.35]{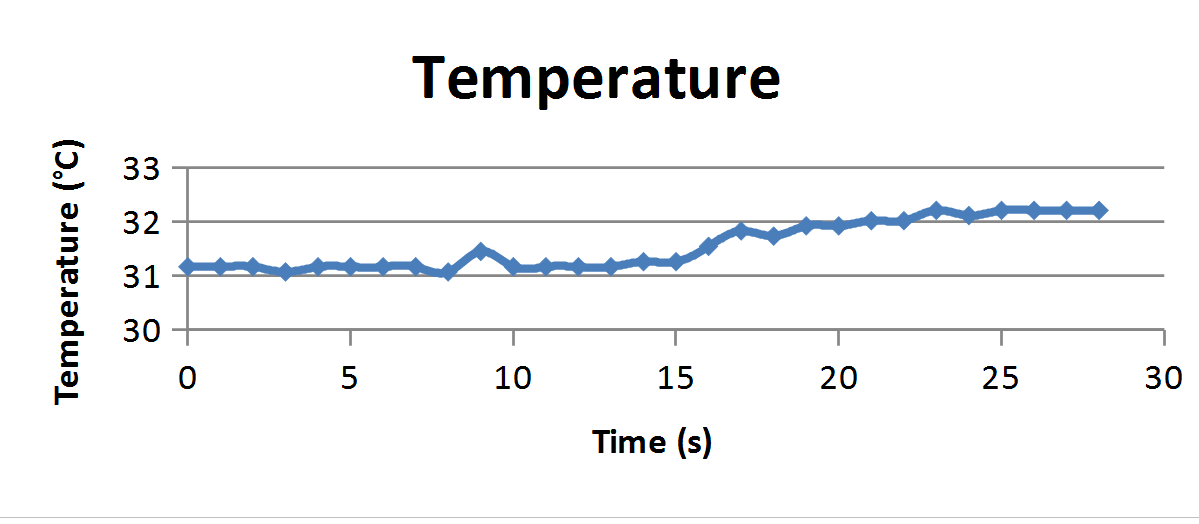}
\caption{Flight test : temperature vs time}\label{flighttemperature}
\end{figure}
The measurements with the temperature sensor showed an excellent accuracy too (Figure~\ref{flighttemperature}). We know that in normal conditions temperature decreases with the increasing of altitude.

The international standard temperature lapse rate is:
\begin{center}
$\dfrac{dT}{dh}=-6.5\dfrac{^{o}C}{1000m}$
\end{center}
They measured the temperature decrease in 165 m :
\begin{center}
$\dfrac{dT}{dh}=-1.04\dfrac{^{o}C}{165m}$
\end{center}
that is:
\begin{center}
$\dfrac{dT}{dh}=-6.3\dfrac{^{o}C}{100m}$
\end{center}
This value is very similar to the international standard value.
\subsubsection{Humidity}
\begin{figure}
\centering
\includegraphics [scale=0.35]{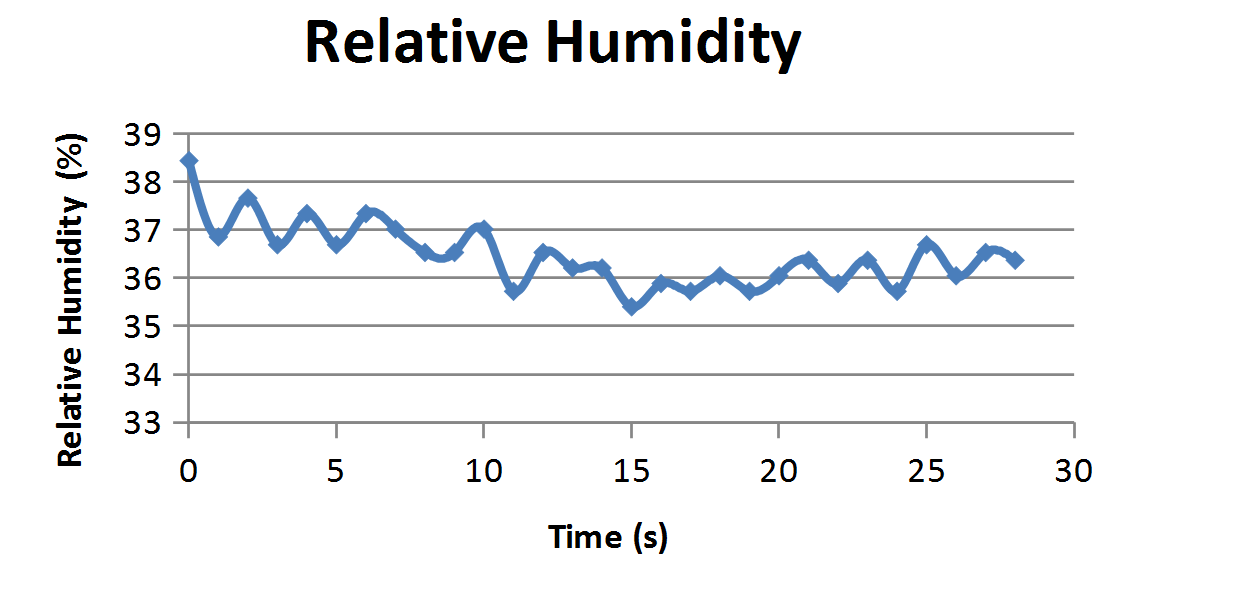}
\caption{Flight test : humidity vs time}\label{flighthumidity}
\end{figure}
Regarding the humidity, students just measured what they expected (Figure~\ref{flighthumidity}) and it was possible to verify a standard relation between relative humidity and temperature.

They obtained that the CanSat at ground level had a temperature of 32.2 $^{o}C$ and relative humidity of 36.4 \% (vapor pressure 36.4 \%  of 4.78 kPa means that vapor pressure is about 1.74 kPa, where 4.78 kPa is the saturation vapor pressure at 32.2 $^{o}C$).

When the CanSat was at the top, the temperature was at minimum value and we can imagine that water vapor remained constant.

The temperature at the top was 31.2 $^{o}C$ and they expected 38.5 \% as theoretical value of relative humidity (4.52 kPa is the saturation vapor pressure at  31.2 $^{o}C$).

The scholars measured a relative humidity of 38.4 \% that is very similar to the one expected.
\subsection{Rocket launch data}
\begin{figure}
\centering
\includegraphics [scale=0.44]{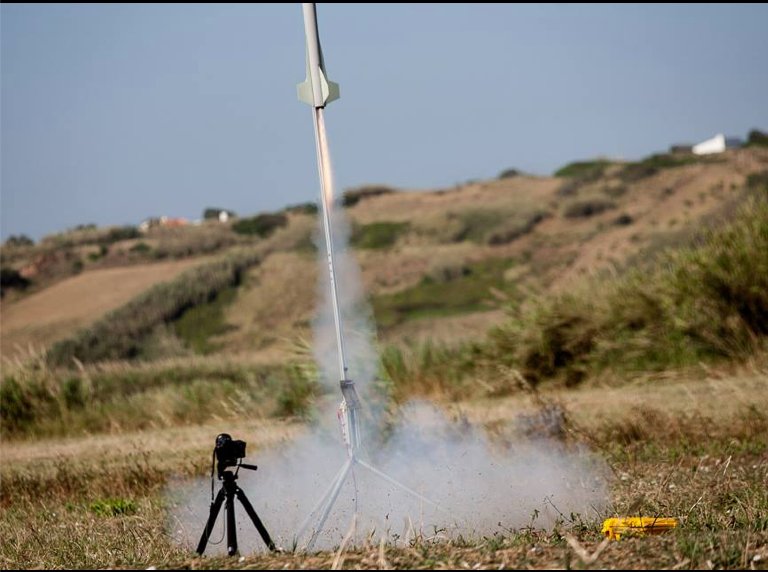}
\caption{CanSat launch}\label{rocket}
\end{figure}

\begin{figure}
\centering
\includegraphics [scale=0.35]{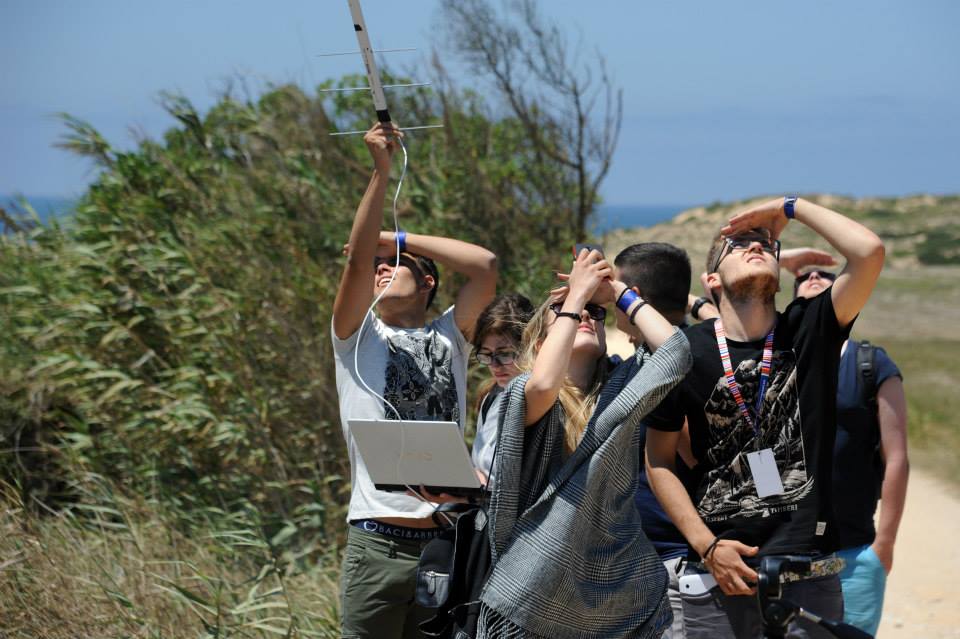}
\caption{Where is my CanSat ?}\label{sun}
\end{figure}

Rocket launch (Figure~\ref{rocket}) was on 26th June 2015 at about 13:00, 2km south of the Santa Cruz Airfield (Portugal ). The cruciform parachute opened correctly and the CanSat + parachute system proceeded with an estimated speed of 6 m/s.
Data transmission was not completely satisfactory:\\
- all data received at ground before launch\\
- only  50\% of data during rocket rise\\
- all data at the top (1076 m altitude estimated using $p_{top}=897hPa$)\\
- 30 \% of data during the descent (Figure~\ref{sun}) of the CanSat\\
- all data at ground after launch.
\subsubsection{Pressure}
\begin{figure}
\centering
\includegraphics [scale=0.40]{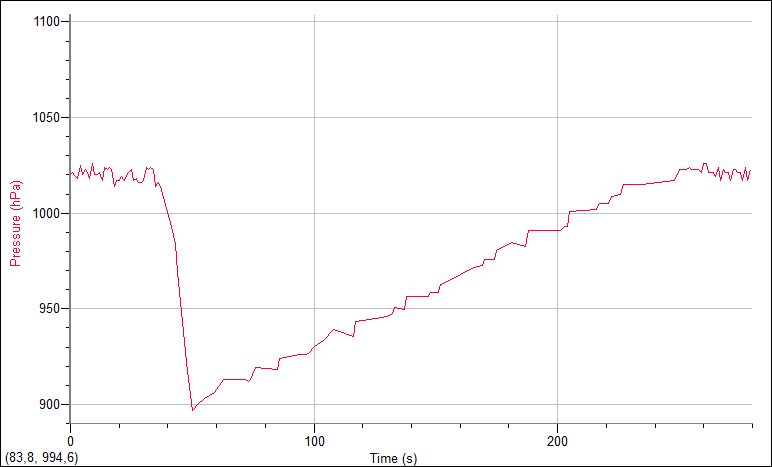}
\caption{Rocket launch : pressure vs time}\label{pressure}
\end{figure}
The pressure-time diagram (Figure~\ref{pressure}) has been obtained using an interpolation: it is important to notice that $t=36s$ is the time in which the CanSat was launched, $t=50s$ is the time in which the rocket arrived at the apogee and $t=235s$ is the landing time (in this diagram there are also the measurements done before and after the launch).

This graph approaches with a good approximation to the one expected, with a sharp decrease between $t=36s$ and $t=50s$  (during the ascent of the rocket) and a gradual rise between $t=50s$ and $t=235s$ (during the descent of the CanSat).

Using the international standard value in the low troposphere:
\begin{center}
$\dfrac{dp}{dh}=-11.15\dfrac{hPa}{100m}$
\end{center}
it is possible calculate 1076 m as apogee.
\subsubsection{Temperature and humidity}
\begin{figure}
\centering
\includegraphics [scale=0.40]{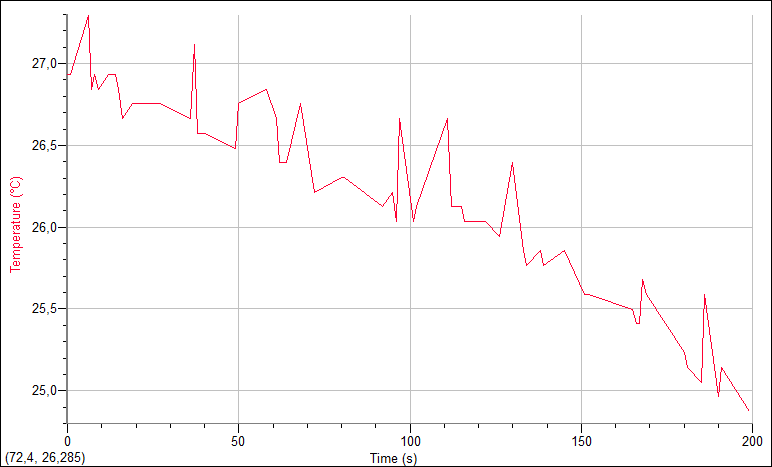}
\caption{Rocket launch : temperature vs time}\label{temperature}
\end{figure}
\begin{figure}
\centering
\includegraphics [scale=0.40]{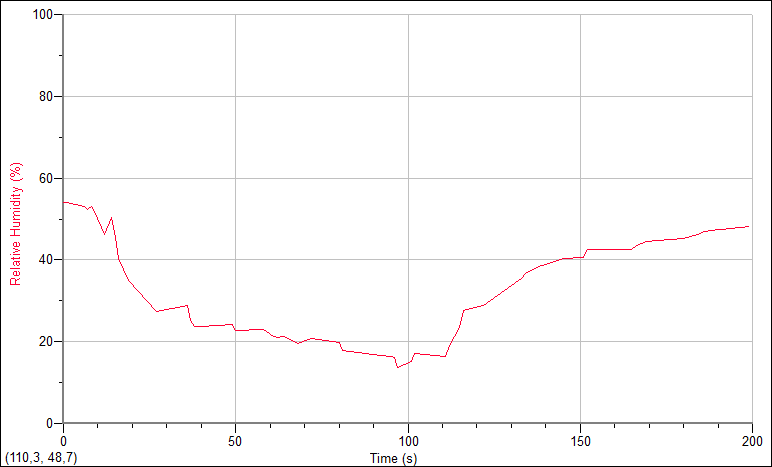}
\caption{Rocket launch : humidity vs time}\label{humidity}
\end{figure}
Using an interpolation pupils obtained the temperature-time diagram (Figure~\ref{temperature}) and the humidity-time diagram (Figure~\ref{humidity}). They considered only the data obtained during the launch. So $t=0s$ is the launch time,  $t=14s$ is the apogee time and $t=199s$ is the landing time.
\section{Conclusions}
My students had no problems to interpret flight test data but on rocket launch they obtained data different from those expected.

In the temperature graph (Figure~\ref{temperature}) during the rise time (between 0 s and 14 s) they expected to measure the maintaining of a constant temperature (or a slight increase). In fact the heating effect caused by the fuel combustion of the rocket and the slow response time of the sensor should have, at least, compensated the expected temperature decrease of 6.5 $^{o}C$ per 1000 m. In this case the forecast was correct: temperature was almost stable at 27 $^{o}C$.

Instead something completely strange occurred during the descent (between $14s$ and $199s$).
They expected a rapid temperature decrease in the immediately following seconds to the release of CanSat (14 s). Then, after reaching a certain minimum temperature, the temperature itself should have began to rise according to the standard temperature lapse rate:
\begin{center}
$\dfrac{dT}{dh}=-6.5\dfrac{^{o}C}{1000m}$
\end{center}
until reaching a maximum on the ground. 
But it was not so. In the first few seconds the temperature started to decline as expected, although actually too slowly, and then continued to decline throughout the descent.

Leaving out the possibility that the sensor failed, first plausible explanation was linked to the holes they made on aluminium CanSat surface. Holes may have generated a current inside the CanSat causing a cooling.

The relative humidity diagram (Figure~\ref{humidity}) shows no correlation with the temperature diagram. A possible reason is that the temperature sensor measured CanSat temperature instead of atmosphere temperature. On the other hand, humidity sensor correctly measured in atmosphere because humidity was the same inside and outside the CanSat.

In fact there is a decrease, then a subsequent increase in relative humidity ( the lowest point is at $t=97s$ ) with a 20 \% variation between the maximum and the minimum.
At $t=97s$ after the launch, pressure was $970 hPa$ that means $421 m$ altitude.

We can imagine a thermal inversion at $421 m$ altitude where a warmer, less-humid parcel of air was over a cooler air mass. This situation usually happens when atmosphere is unstable and convection is likely, in fact the CanSat was launched at 1 p.m. with about 30 $^{o}C$ at ground level during a period of strong insolation.
\section*{Acknowledgements}
Special thanks to my students Augusto Basilico, Riccardo Colleluori, Mattia Davario, Aurora Forcella, Ilaria Pavone, Simone Piovani. They won the third prize in the beginners category of the 2015 European CanSat competition.
\bibliography{arxiv}{}
\bibliographystyle{plain}
\section*{Appendix CanSat Requirements}
The CanSat hardware and missions must be designed to the following requirements and constraints:

[1]	All the components of the CanSat must fit inside a standard soda can (115 mm height and 66 mm diameter), with the exception of the parachute. An exemption can be made for radio antennas and GPS antennas, which can be mounted externally (on the top or bottom of the can, not on the sides), based on the design.
	N.B. The rocket payload area has 4.5 cm of space available per CanSat, along the can’s axial dimension (i.e. height), which must accommodate all external elements including: parachute, parachute attachment hardware, and any antennas.\\

[2]	The antennas, transducers and other elements of the CanSat cannot extend beyond the can’s diameter until it has left the launch vehicle.\\

[3]	The mass of the CanSat must be between 300 grams and 350 grams. CanSats that are lighter must take additional ballast with them to reach the 300 grams minimum mass limit required.\\

[4]	Explosives, detonators, pyrotechnics, and flammable or dangerous materials are strictly forbidden. All materials used must be safe for the personnel, the equipment and the environment. Material Safety Data Sheets (MSDS) will be requested in case of doubt.\\

[5]	The CanSat must be powered by a battery and/or solar panels. It must be possible for the systems to be switched on for four continuous hours.\\

[6]	The battery must be easily accessible in case it has to be replaced/recharged.\\

[7]	The CanSat must have an easily accessible master power switch.\\

[8]	Inclusion of a retrieval system (beeper, radio beacon, GPS, etc.) is recommended.\\

[9]	The CanSat should have a recovery system, such as a parachute, capable of being reused after launch. It is recommended to use bright coloured fabric, which will facilitate recovery of the CanSat after landing.\\

[10]	The parachute connection must be able to withstand up to 1000 N of force. The strength of the parachute must be tested, to give confidence that the system will operate nominally.\\

[11]	For recovery reasons, a maximum flight time of 120 seconds is recommended. If attempting a directed landing then a maximum of 170 seconds flight time is recommended.\\

[12]	A descent rate between 8 m/s and 11 m/s is recommended for recovery reasons. In case of attempting a directed landing, a lower descent rate of 6m/s is recommended. \\

[13]	The CanSat must be able to withstand an acceleration of up to 20 g.\\

[14]	The total budget of the final CanSat model should not exceed: 500\euro for the Beginners category and 600\euro for the Advanced category. Ground Stations (GS) and any related non-flying item will not be considered in the budget. More information regarding the penalties in case of exceeding the stated budget can be found in the next section.\\

[15]	In case of sponsorship, all the items obtained should be specified in the budget with the corresponding costs on the market at that moment.\\

[16]	The CanSat must be flight-ready upon arrival. A final technical inspection of the CanSats will be done by authorised personnel before launch.

\end{document}